\newcommand{\nc}{\newcommand}
\nc{\be}{\begin{equation}}
\nc{\ee}{\end{equation}}
\nc{\bea}{\begin{eqnarray}}
\nc{\eea}{\end{eqnarray}}
\nc{\beas}{\begin{eqnarray*}}
\nc{\eeas}{\end{eqnarray*}}
\nc{\ra}{\rightarrow}
\nc{\s}[1]{\not \! \!  #1}
\nc{\lapp}{\hbox{$ {     \lower.40ex\hbox{$<$}
                   \atop \raise.20ex\hbox{$\sim$}
                   }     $}  }
\nc{\rapp}{\hbox{$ {     \lower.40ex\hbox{$>$}
                   \atop \raise.20ex\hbox{$\sim$}
                   }     $}  }

\documentstyle{aipproc}

\begin{document}
\rightline{ UK/TP 97-17}
\rightline{ ADP-97-27/T262}
\rightline{ hep-ph/9707414}

\title{Direct CP Violation in 
${\bf B^{\pm}\rightarrow\rho^{\pm}\pi^+\pi^-}$ in the 
$\rho^0$-$\omega$ Interference Region \thanks{
Talk presented by S. Gardner at the $6^{\rm th}$ Conference on the
Intersections of Particle and Nuclear Physics, Big Sky, MT,
May 27 - June 2, 1997.}}

\author{S. Gardner, H.B. O'Connell$^*$, and A.W. Thomas$^{\dagger}$}
\address{$^*$Department of Physics and Astronomy, \\ University of Kentucky, 
        Lexington, KY 40506-0055 USA \\
$^{\dagger}$Department of Physics and Mathematical Physics, \\ and 
Special Research Centre for
the Subatomic Structure of Matter, \\
University of Adelaide, Adelaide, S.A. 5005 AUSTRALIA}

\maketitle

\begin{abstract}
We study direct CP violation in 
$B^{\pm}\rightarrow\rho^{\pm}\rho^0(\omega)\rightarrow\rho^{\pm}\pi^+\pi^-$
and focus specifically on the rate asymmetry in the $\rho^0$-$\omega$
interference region. Here the
strong phase is dominated by isospin violation, so that
it can be essentially determined 
by $e^+e^- \rightarrow \rho^0 (\omega) \rightarrow \pi^+ \pi^-$ data. 
We find the CP-violating asymmetry to be of the order of 20\%
at the $\omega$ invariant mass. Moreover, it is robust 
with respect to the estimable strong-phase
uncertainties, permitting the extraction of
$\sin\alpha$ from this channel. 
\end{abstract}

  Experimental programs in the next years at HERA, KEK, and SLAC 
will study CP violation in the $B$-meson system in the hope of identifying 
physics beyond the Standard Model. 
In the Standard Model, the so-called unitarity triangle associated with the
CKM parameters $\alpha$, $\beta$, and $\gamma$ requires that these
angles sum to $\pi$ [1]. Yet, the experiments in 
the neutral B sector which would measure $\alpha$, {\it e.g.}, 
determine merely $\sin 2 \alpha$ [2], 
so that discrete ambiguities remain in 
$\alpha$ itself [3]. Here we consider CP violation in 
$B^\pm \rightarrow \rho^\pm \rho^0(\omega) \rightarrow \rho^\pm \pi^+ \pi^-$,
with $\rho^0 (\omega)$ denoting the $\rho^0$-$\omega$  interference
region, proposed by Enomoto and Tanabashi [4]. 
The rate asymmetry, which is CP-violating, arises exclusively from a 
nonzero phase in the CKM matrix, so that the CP violation is
termed ``direct.'' The manner in which CP-violation is generated
differs from that of the neutral meson case, so that the asymmetry
depends on $\sin \alpha$. Its determination, then, removes the 
${\rm mod}(\pi)$ ambiguity in $\alpha$ inherent in the $\sin 2\alpha$
measurement [3]. The rub, however, is that direct CP violation 
requires that both a strong and weak phase difference exist between
two interfering amplitudes [5], so that information on the 
weak phase is generally obscured by 
strong interaction uncertainties. In the above case, however, 
data in $e^+e^- \rightarrow \pi^+ \pi^-$ in the 
$\rho^0$-$\omega$ inteference region substantially constrains the
strong phase [6]. 
Here we discuss the remaining 
uncertainties in the $\sin\alpha$ determination, and show how they may 
be obviated with further experimental data. 

Resonances can play a strategic role 
in direct CP violation. Resonance information, such as the mass and
width, can be used 
to constrain the strong phase, and their interference can significantly
enhance the CP-violating asymmetry [7,8]. 
Both these effects are operative in hadronic B-decays in the 
$\rho^0$-$\omega$ interference region [4,9]. In 
$B^{\pm}\ra \rho^{\pm}\rho^0(\omega)$, 
$\rho$-$\omega$ interference is the dominant source of strong phase, and 
it can be determined through
fits to $e^+e^- \rightarrow \pi^+ \pi^-$ data [6].
We also have computed the additional isospin violating effects arising from 
electroweak penguin contributions and within the $\rho^-$ and $\rho^0$ 
hadronic form factors. Their impact is small, however, relative to that of
the $\rho$-$\omega$ mixing contribution. 
The asymmetry we predict is of order of 
20\% at the $\omega$ invariant mass; the asymmetry is both 
large and robust with respect to the known strong phase uncertainties. 

The CP-violating asymmetry in 
$B^{\pm}\rightarrow \rho^{\pm}\pi^+ \pi^-$
 is significantly enhanced by $\rho^0$-$\omega$ 
mixing. To see why this is so, consider the amplitude 
$A$ for $B^- \rightarrow \rho^- \pi^+ \pi^-$ decay:
\be
A = \langle \pi^+\pi^- \rho^- | {\cal H}^{\rm T} | B^- \rangle
+ \langle \pi^+\pi^- \rho^- | {\cal H}^{\rm P} | B^- \rangle \;,
\ee
where $A$ is given by the sum of the amplitudes corresponding to the tree
and penguin diagrams, respectively. Defining the strong phase
$\delta$, the weak phase $\phi$, and the magnitude $r$ via 
\be
A = \langle \pi^+\pi^- \rho^- | {\cal H}^{\rm T} | B^- \rangle \left[
1 + re^{i\delta}\; e^{i\phi} \right] \;,
\ee
one has 
$\overline{A}
=\langle \pi^+\pi^- \rho^+ | {\cal H}^{\rm T} | B^+ \rangle 
[1 + re^{i\delta}\; e^{-i\phi}]$.
Thus, the CP-violating 
asymmetry $A_{{\s{\rm CP}}}$ is 
\be
A_{{\s{\rm CP}}} \equiv 
{ | A |^2 - |{\overline A}|^2 
\over | A |^2 + |{\overline A}|^2 }
= {-2r \sin \delta \sin \phi 
\over 1 + 2r\cos \delta \cos \phi + r^2 } \;,
\label{asym}
\ee
so that both $\delta$ and $\phi$ must be non-zero to yield a non-zero
asymmetry. Here $\phi$ is $- \alpha$ [1].

To express $\delta$ in terms of the resonance
parameters, 
let $t_{\rm V}$ be the tree amplitude 
and $p_{\rm V}$ be the penguin amplitude to produce a vector meson ${\rm V}$. 
Thus, the tree and penguin 
amplitudes for $B^-\rightarrow \rho^- \pi^+ \pi^-$ can be written as 
\bea
\langle \pi^+\pi^- \rho^- | {\cal H}^{\rm T} | B^- \rangle 
= {g_{\rho} \over s_\rho s_\omega} \tilde\Pi_{\rho\omega} t_{\omega}
  + { g_{\rho} \over s_\rho } t_\rho  \;,\\
\langle \pi^+\pi^- \rho^- | {\cal H}^{\rm P} | B^- \rangle 
= {g_{\rho} \over s_\rho s_\omega} \tilde\Pi_{\rho\omega} p_{\omega}
  + { g_{\rho} \over s_\rho } p_\rho \;. \label{fun}
\eea
Note that $\tilde\Pi_{\rho\omega}$ is the effective 
$\rho^0$-$\omega$ mixing matrix
element, 
 $g_\rho$ is the $\rho^0 \rightarrow \pi^+\pi^-$ coupling, $1/s_{\rm V}$
is the vector meson propagator, 
$s_{\rm V}=s - m_{\rm V}^2 + i m_{\rm V} \Gamma_{\rm V}$, 
with $m_{\rm V}$ and $\Gamma_{\rm V}$ the vector meson mass and width, 
and $s$ is the invariant mass of the $\pi^+ \pi^-$ pair. 
$\tilde\Pi_{\rho\omega}$ is extracted from
pion form-factor data, as measured in 
$e^+e^-\rightarrow \pi^+\pi^-$, and it 
is insensitive to the ambiguities in the $\rho$ 
parametrization [10]. We have fit 
$\tilde\Pi_{\rho\omega}(s)=\tilde\Pi_{\rho\omega}(m_\omega^2) 
+ (s - m_\omega^2)\tilde\Pi_{\rho\omega}'(m_\omega^2)$ to find
$\tilde\Pi_{\rho\omega}(m_\omega^2) = -3500\pm 300$ MeV$^2$ and 
$\tilde\Pi_{\rho\omega}'(m_\omega^2) = 0.03 \pm 0.04$ [10]. 
Using 
\be
re^{i\delta}\,e^{i\phi}= 
{ \langle \pi^+\pi^- \rho^- | {\cal H}^{\rm P} | B^- \rangle 
\over 
\langle \pi^+\pi^- \rho^- | {\cal H}^{\rm T} | B^- \rangle}
= 
{ \tilde\Pi_{\rho\omega} p_{\omega} + s_\omega p_\rho
\over
\tilde\Pi_{\rho\omega} t_{\omega} + s_\omega t_\rho} \;,
\ee
and the definitions of Ref. [4]:
\be
{p_\omega \over t_\rho} \equiv r' e^{i(\delta_q + \phi)} \;, \quad
{t_\omega \over t_\rho} \equiv \alpha e^{i \delta_\alpha} \;, \quad
{p_\rho \over p_\omega} \equiv \beta e^{i \delta_\beta} \;,
\label{alphabeta}
\ee
one finds, to leading order in isospin violation, 
\be
re^{i\delta} = 
{r' e^{i\delta_q}\over s_\omega} \left\{
 \tilde\Pi_{\rho\omega} + \beta e^{i\delta_\beta}
\left( s_\omega 
- \tilde\Pi_{\rho\omega} \alpha e^{i\delta_\alpha} \right)
\right\} 
\;.
\label{thescoop}
\ee
Note that $\delta_\alpha,\delta_\beta,\hbox{and}\;\delta_q$ are 
``short-distance'' phases, 
generated by 
putting the quarks in loops
on their mass-shell [5]; this mechanism is the typical 
source of strong phase. 
A $J=0,\,I=0$ $\rho^\pm\rho^0$ 
final state is forbidden by Bose symmetry if isospin is perfect, 
so that
$\beta$ is non-zero only if electroweak penguin contributions and
isospin violation in the $\rho^\pm$ and $\rho^0$ hadronic form
factors are included. Numerically, 
$|\tilde\Pi_{\rho\omega}|/(m_\omega \Gamma_\omega) \gg \beta$.
The resonant enhancement
of the CP-violating asymmetry is thus driven by 
$\tilde\Pi_{\rho\omega}/s_\omega$.
As $s\rightarrow m_\omega^2$, the asymmetry is maximized if 
$|\chi| \equiv |\tilde\Pi_{\rho\omega}|/m_\omega\Gamma_\omega\sim O(1)$ and
$\delta_q + \eta \sim \pm\pi/2$, where 
$\eta=-{\rm arg}\; s_\omega$. 
Here $|\chi|\approx .53$ [10,1] and $\eta= -\pi/2$. 
Note that $\delta_q \lapp -161^\circ$ [4], so
that $\delta_q +\eta \lapp -251^\circ$ at the $\omega$ mass.

The CP-violating asymmetry from Eqs.~(\ref{asym},\ref{thescoop}), then, is 
determined by the resonance parameters
$\tilde\Pi_{\rho\omega}$, $m_\omega$, $\Gamma_\omega$, and the
``short distance'' parameters $\alpha$, $\delta_\alpha$, $\beta$,
$\delta_\beta$, $r'$, $\delta_q$, as well as  $\phi$, the weak phase. 
The latter class of 
parameters are calculable within the context of the 
operator product expansion if the factorization approximation is
applied, though a ratio of hadronic form factors
enters as well. Here that ratio is modified from unity by 
isospin-violating effects only. Using the above method, we find
asymmetries of the order of 20\% at the $\omega$ invariant mass,
and an asymmetry of this magnitude is retained even if the imaginary parts
of the effective Wilson coefficients are set to zero [6]. 
 
From our earlier discussion, however, it is clear that the
sign of $\sin\phi$ is of unique significance, so that it is useful
to consider how it may be extracted. If $r < .5$ then the sign of the
CP-violating asymmetry is determined by $\sin\delta$ and $\sin\phi$. 
As $s\ra m_\omega^2$, the sign of $\sin\delta$ is determined by 
\be
{\rm sgn}(\sin\delta) 
= {\rm sgn}( 
\cos\delta_q \,{\rm Im} \Omega + \sin\delta_q \,{\rm Re} \Omega) \;,
\ee
where 
$\Omega =\beta(\cos\delta_\beta - \chi \sin\delta_\beta)
- i (\chi (1 - \beta\cos\delta_\beta) - \beta\sin\delta_\beta)$
in this limit, recalling 
$\chi=\tilde\Pi_{\rho\omega}/(m_\omega \Gamma_\omega)$. 
As $|\chi| > \beta$, the sign of $\sin\delta$ is
determined by $-\chi \cos\delta_q$. The sign of $\chi$ is just that
of $\tilde\Pi_{\rho\omega}$, but what
 of that of $\cos\delta_q$? To determine this, note that the 
``skew'' of the asymmetry --- the sign of the $(s - m_\omega^2)$ term 
multiplying $1/|s_\omega|$ 
in $\sin\delta$ from Eq.~(\ref{thescoop}) ---
determines the sign of $\sin\delta_q$ as $|\chi| > \beta$. 
Note that the empirical
$s$-dependence of $\tilde\Pi_{\rho\omega}(s)$ about 
$s=m_\omega^2$ does not cloud this interpretation [6]. 
In the factorization approximation, 
the sign of $\sin\delta_q$ is invariably that of $\cos\delta_q$ [4], 
yet one need not assume this. The sign of the asymmetry in 
$B^\pm\ra \rho^\pm \omega \ra \rho^\pm \pi^+ \pi^- \pi^0$ 
is driven by $\sin\delta_q\sin\phi$, and it is also large [11]. 
The signs of
the two asymmetries, whether they are the same or different, determines
the sign of $\cos\delta_q$ once the sign of $\sin\delta_q$ is
known. In this manner, the sign of $\sin\phi$, or specifically
that of $-\sin\alpha$ [1], is determined without the need of the
factorization approximation. The assumptions needed are that $r <.5$ and
$|\chi| > \beta$ ---  both are borne out in our analysis [6]. 
The data needed in order to effect this extraction are the asymmetry and its 
shape in $B^\pm \ra \rho^\pm \rho^0 (\omega) \ra \rho^\pm \pi^+ \pi^-$ 
about $s=m_\omega^2$ and the asymmetry in 
$B^\pm \ra \rho^\pm \omega \ra \rho^\pm \pi^+ \pi^- \pi^0$. 

We thank H.J. Lipkin for helpful discussions, and A. Kagan, W. Korsch, and
G. Valencia for useful comments and references.


\begin{references}

\bibitem{1} R.M. Barnett {\it et al.}, 
{\it Phys. Rev. D}\ {\bf 54}, 1 (1996). 

\bibitem{2} M. Gronau and D. London, 
{\it Phys. Rev. Lett.}\ {\bf 65}, 3381 (1990); A.E. Snyder and H.R. Quinn,
{\it Phys. Rev. D}\ {\bf 48}, 2139 (1993). 

\bibitem{3}  Y. Grossman and H.R. Quinn, hep-ph/9705356. 

\bibitem{4} R. Enomoto and M. Tanabashi, 
{\it Phys. Lett. B}\ {\bf 386}, 413 (1996); see also hep-ph/9706340. 

\bibitem{5} M. Bander, D. Silverman, and A. Soni, 
{\it Phys. Rev. Lett.}\ {\bf 43}, 242 (1979). 

\bibitem{6} S. Gardner, H.B. O'Connell, and A.W. Thomas, hep-ph/9705453.

\bibitem{7} D. Atwood and A. Soni, 
{\it Phys. Rev. Lett.}\ {\bf 74}, 220 (1995); 
{\it Z. Phys.~C} {\bf 64}, 241 (1994). 

\bibitem{8} D. Atwood, G. Eilam, M. Gronau, A. Soni, 
{\it Phys. Lett.~B}\ {\bf 341}, 372 (1995); 
G. Eilam, M. Gronau, and 
R.R. Mendel, {\it Phys. Rev. Lett.}\ {\bf 74}, 4984 (1995). 

\bibitem{9} H.J. Lipkin, hep-ph/9310318; {\it Phys. Lett. 
B}\ {\bf 357}, 404 (1995). 

\bibitem{10} S. Gardner and H.B. O'Connell, hep-ph/9707385.

\bibitem{11} S. Gardner, H.B. O'Connell, and A.W. Thomas, in preparation.

\end{references}
\end{document}